\documentclass{ws-procs961x669}            

\usepackage{hyperref}

\begin{document}

\title{General Relativity and Geodesy}

\author{E. Hackmann$^1$, C. L\"ammerzahl$^2$, D. Philipp$^3$, and B. Rievers$^4$}

\address{Center of Applied Space Technology and Microgravity (ZARM), University of Bremen,\\
28359 Bremen, Germany and\\
Gau{\ss}-Olbers Space Technology Transfer Center (GOC), University of Bremen \\
28359 Bremen, Germany\\
$^1$E-mail: eva.hackmann@zarm.uni-bremen.de\\
$^2$E-mail: claus.laemmerzahl@zarm.uni-bremen.de\\
$^3$E-mail: dennis.philipp@zarm.uni-bremen.de\\
$^4$E-mail: benny.rievers@zarm.uni-bremen.de\\
www.zarm.uni-bremen.de}

\author{M. Huckfeldt}

\address{Center of Applied Space Technology and Microgravity (ZARM), University of Bremen,\\
28359 Bremen, Germany\\
E-mail: moritz.huckfeldt@zarm.uni-bremen.de}

\begin{abstract}
Mass redistribution on Earth due to dynamic processes such as ice melting and sea level rise leads to a changing gravitational field, observable by geodetic techniques. 
Monitoring this change over time allows us to learn more about our planet and its dynamic evolution.
In this paper, we highlight the impact of General Relativity (GR) on geodesy: it provides corrections essential for the interpretation of high-precision measurements and enables a completely novel measurement approach using chronometry, i.e., clock-based observations. 
Focusing on the latter, we review the construction of the relativistic gravity potential and the corresponding geoid definition as an isochronometric surface to elucidate the comparison to the conventional Newtonian geoid. 
Furthermore, we comment on additional potentials due to the non-Newtonian degrees of freedom of the relativistic gravitational field, and assess the feasibility of clock-based measurements for Gravity Field Recovery (GFR) from space. 
Although clock observations in space demonstrate technical promise for GFR, achieving the necessary precision for practical applications remains challenging. 
\end{abstract}


\bodymatter


\section{Introduction}

\begin{figure}
\centering
\includegraphics[width=0.9\textwidth]{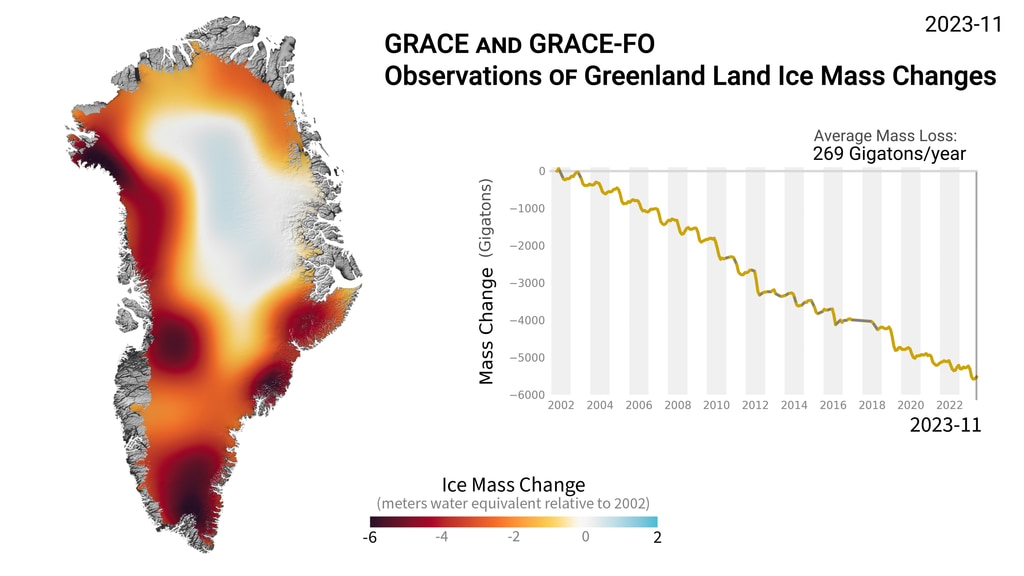}
\caption{Changes in Greenland ice mass as observed by the GRACE and GRACE-FO satellites, using the monthly surface mass anomaly data from 2002 to 2023. The average mass loss of 269 gigatons per year can also be expressed as about 0.8 times the mass of the Eiffel tower per second. Credit: NASA and JPL/Caltech.}
\label{fig:greenland}
\end{figure}

Several dynamic processes in the System Earth are related to the redistribution of masses. 
For example, due to climate change, we are facing floodings, ice melting, land uplift, and sea level rise as well as droughts and irrigation. 
Because such processes lead to a change of the mass-energy distribution, they can be observed by monitoring the gravitational field of the Earth and its change over time within the field of geodesy. 
One geodetic method to realize this is satellite gravimetry using the Gravity Recovery and Climate Experiment (GRACE) and GRACE Follow-On (GRACE-FO) satellite missions. 
In figure \ref{fig:greenland} an example is shown to demonstrate how time-resolved gravity field data can give insight into ice melting processes in Greenland. 
The clear, linear trend shows that on average there is a daily ice mass loss, which corresponds to a land area of $1\,$km$^2$ covered by ice of more than $700\,$ meters in height, assuming average ice density.
Other methods to observe the gravity field include, but are not limited to, local measurements of the gravitational acceleration with quantum gravimeters, or GNSS station displacements. 

\begin{figure}
    \centering
    \includegraphics[width=0.4\linewidth]{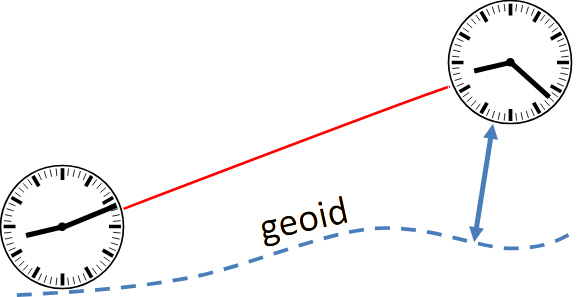}\qquad 
    \includegraphics[width=0.4\linewidth]{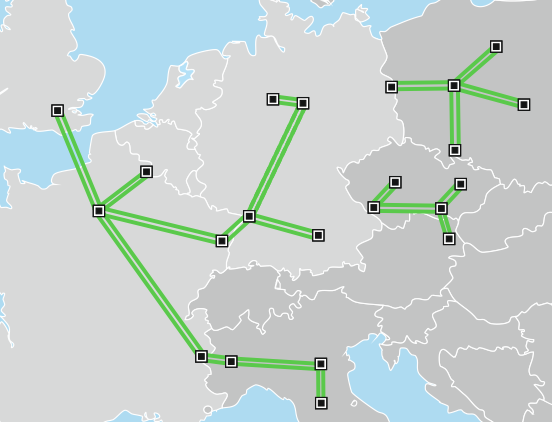}
    \caption{Left: Schematic illustration of the concept of chronometric levelling. The gravitational redshift is related to the difference in gravitational potential at the two clock sites, and can therefore be translated into a height measure. Right: Illustration of high quality optical fibre links in Europe. Credit: PTB.}
    \label{fig:redshift-lev}
\end{figure}

In this review, we focus on the role of General Relativity (GR) in geodesy. 
On the one hand, GR sets the stage for all physics and measurements in spacetime. 
With respect to the increasing measurement accuracies, relativistic effects, e.g., on satellite motion, must be consistently included. 
On the other hand, in GR gravity is geometry. 
As clocks directly probe the spacetime geometry by their proper time, a completely new measurement approach for the gravitational field arises. 
The gravitational redshift was in the context of geodesy first considered by Vermeer \cite{vermeer_chronometric_1983} and Bjerhammer\cite{Bjerhammar1985}. To lowest post-Newtonian order, it is given by
\begin{align}
z = \frac{\Delta \nu}{\nu_2} = \frac{\Delta U}{c^2} + \mathcal{O}\big(c^{-4}\big)
\end{align}
where $\Delta \nu := \nu_1 - \nu_2$ are the frequencies, $\Delta U := U_1 - U_2$ is the gravitational potential differences, and $c$ is the speed of light. 
See also figure \ref{fig:redshift-lev} on the left for an illustration of the general idea.

It is assumed that the two clocks that measure this redshift have no relative motion, i.e., they rest on the Earth's surface.
From this simple formula one can estimate that for a height resolution with centimeter accuracy, close to the surface, clocks with uncertainties in the $10^{-18}$ regime are needed, which is within reach. 
Additional benefits in using optical clocks for height determination are high spatial resolution, as atoms are small, no error accumulation over distance, and fast measurements.
In addition to two (transportable) clocks on the $10^{-18}$ level, a link for the comparison is needed that is good enough to not spoil the clock accuracy.

There have been first experimental campaigns to realize height determination via clock comparisons, called chronometric levelling. 
In the framework of the TerraQ collaboration\footnote{see https://www.terraq.uni-hannover.de/en/}, the most stable transportable clock lasers have been developed \cite{Herbers:22} and been used to realize a transportable Strontium lattice clock with an uncertainty below $3 \times 10^{-18}$, see Refs.~\citenum{Lisdat2021} and \citenum{Doerscher2023}. 
Moreover, high quality optical fibre links have been established, see figure \ref{fig:redshift-lev} on the right, with instabilities below $10^{-18}$ in $100\,$s integration time and offsets below $10^{-19}$, see Refs \citenum{Schioppo2022,Koke:2019}.
In addition to the fibre network, GNSS frequency transfer has been demonstrated with instabilities below $5 \times 10^{-17}$ on a 50 km baseline\cite{Elmaghraby2023}. On a  short distance of a few hunderd meters, chronometric levelling has been demonstrated on the centimeter level in Japan \cite{Takamoto2020}. Long distance chronometric levelling has been accomplished with accuracies on the  decimeter level \cite{Grotti2024} in a measurement campaign between Munich and Braunschweig in Germany, that is currently repeated with better clocks.

In this review we focus on the general relativistic foundational notions in geodesy, discuss the potential for geodesic applications, and comment on recent investigations on applications involving clocks in space.

\section{General relativistic framework for geodesy}	

Important aspects of geodesy in the context of gravitational field observations include the (post-)Newtonian gravity potential, reference surfaces and systems, as well as height definitions. 
While post-Newtonian treatments of these concepts have been developed some time ago, see e.g., Refs.\ \citenum{muller2008geodesy,Soffel:2003cr} and references therein, a consistent definition of such geodetic concepts in the framework of the full theory of General Relativity started only in recent years. 
We will review some of these developments here, show how they shed light on subtleties, and reproduce the known post-Newtonian expressions.  

Let us begin with the definition of fundamental potentials in conventional geodesy. 
In the Newtonian theory, the gravity potential (geopotential) is
\begin{align}
	\label{Eq_NewtonianTotalPotential}
	W = U + V \,,
\end{align}
where $V$ is the centrifugal part and $U$ is the gravitational potential. 
Here we deliberately distinguish between gravity and gravitation as is commonly done in geodesy. 
The gravitational potential is usually expanded into spherical harmonics,
\begin{align}
	\label{Eq_NewtonianPotenialDecomposition}
	U = \dfrac{GM}{r} \sum_{l=0}^\infty \sum_{m=0}^l \left( \dfrac{R}{r} \right)^{l} P_{lm}(\sin \xi) \left\lbrace C_{lm} \cos(m \lambda)  +  S_{lm} \sin(m \lambda) \right\rbrace \, ,
\end{align}
where $M$ is the mass of the Earth in kg, $R$ is some arbitrary reference radius (e.g. the equatorial radius) in meters, $r, \xi, \lambda$ are radius, latitude and longitude in geocentric spherical coordinates, $l, m$ are the degree and order of the expansion, $P_{l,m}$ are the associated Legendre functions, and $C_{lm}, S_{l,m}$ are the multipole coefficients of the respective expansion degree and order.

The gravity potential $W$ foliates three-dimensional Euclidean space into equipotential surfaces. 
One of these, historically meant to be as close to mean sea level as possible, is termed the geoid. 
Today, the geoid is fixed by convention to the value $W|_{\text{geoid}}=W_0 = 6.26368534 \times 10^7 \, \rm{m^2/s^2}$, see, e.g., Refs.\ \citenum{TorgeMueller+2012,sanchez2016conventional}.
Note that in geodesy several different versions of "geoids" exist that differ in the assumptions included in the definition, like the inclusion of (permanent or non-permanent) tidal effects. 
Here, we will only consider the geoid in the absence of any external forces, assume a constant rotation axis, and constant angular velocity of the Earth. 

The above definition of the geoid is realized by measuring accelerations that yield local plumb lines. 
The equipotential surfaces of $W$ are perpendicular to the plumb lines and are then transported in levelling networks.
Due to the measurement scheme, this geoid is sometimes called the a-geoid\cite{soffel1988relativistic}.

The most prominent effect of GR in the context of geodesy is the gravitational redshift, which relates potential differences to frequency shifts. 
A corresponding definition of a relativistic geoid was given (just in words) by Bjerhammer\cite{Bjerhammar1985} as the \textit{surface where precise clocks run with the same speed}, so as a surface on which distributed standard clocks (atomic clocks) show a vanishing mutual redshift. 
In this definition, the geoid is based on frequency measurements and sometimes called the u-geoid. 
Here, u refers to the usual symbol used for the four-velocity in relativistic physics\cite{soffel1988relativistic}.

A post-Newtonian realization of the geoid is defined in Ref.\ \citenum{soffel1988relativistic}, and the general relativistic geoid undulation in Ref.\ \citenum{kopeikin2015towards}. 
More recently, a general relativistic definition of the geoid is given in Ref.\ \citenum{oltean2016geoids}, following in spirit the a-geoid approach in arbitrary spacetimes, and in Ref.\ \citenum{philipp2017definition}, based on the u-geoid approach in stationary spacetime, but including comparisons through fiber links as in the above mentioned measurement campaigns. We review here in some more detail the u-geoid approach.

\subsection{Relativistic gravity potential and the geoid}

\begin{figure}
\centering
\includegraphics[width=0.3\textwidth]{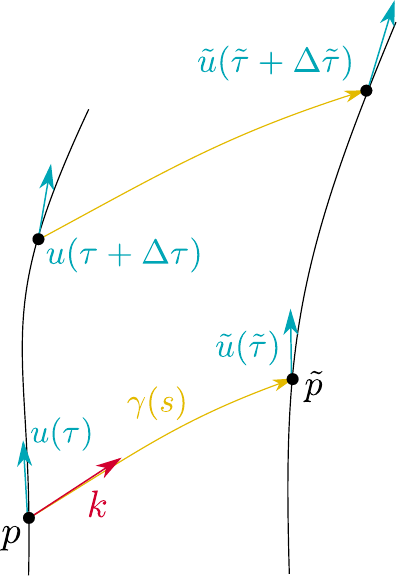}
\caption{Illustration for the definition of the redshift between two timelike worldline on which events are connected by light rays.}
\label{fig:redshift}
\end{figure}

Consider two standard clocks that measure proper times $\tau_1$ and $\tau_2$ on worldlines with four velocities $u_1$ and $u_2$, respectively. 
Assume that a light ray with a tangent covector $k$ is emitted at an event $p$ by the first clock and received at an event $\tilde{p}$ by the second clock, see Fig. \ref{fig:redshift}, and that this connecting light ray is unique.
The redshift between the two clocks connected via events $p$ and $\tilde{p}$ is then given by 
\begin{align}
	\label{Eq_RedshiftDefinition2}
z+1  := \frac{d \tau_2}{d \tau_1}  = \frac{\nu_1}{\nu_2} = \frac{\left. \left( g_{\mu\nu} k^\mu u_1^\nu \right) \right|_{p} }{\left. \left( g_{\rho\sigma} k^\rho u_2^\sigma \right) \right|_{\tilde{p}}} \, ,	
\end{align}
see also the first derivation of this result in Ref.\ \citenum{Kermack1934}.
Here, $g_{\mu\nu}$ is the spacetime metric.
Note that the equation above contains kinematic and gravitational contributions to the total redshift.

In stationary spacetimes, to describe the redshift effect for special observers, we can introduce a (dimensionless) \textit{redshift potential} $\phi$. 
For this, consider a family of observers described by a congruence of worldlines with a tangent vector field $u$. 
If and only if $\exp(\phi)u =: \xi$ is a Killing vector field of the spacetime, there exists a time-independent redshift potential $\phi$ with the property
\begin{align}
	\label{Eq_RedshiftPotentialDefinition}
	\log (z+1) = \phi|_{\tilde{\gamma}} - \phi|_{\gamma}\,,
\end{align}
for any two clock worldlines $\gamma, \tilde{\gamma}$ in the congruence, see Ref.\ \citenum{HassePerlick1988}. 
We call surfaces of constant redshift potential \textit{isochronometric surfaces}. 
All standard clocks $\gamma_i, \, i=1 \dots N$ that are on the same surface $\phi = \phi_0 = \text{const.}$ show zero redshift w.r.t.\ each other.

An important remark is that although in Eq.\ \eqref{Eq_RedshiftDefinition2} it is assumed that the light signal propagates along lightlike geodesics, it was proven in Ref \citenum{philipp2017definition} that \eqref{Eq_RedshiftPotentialDefinition} also holds true if the signal propagates along an arbitrary optical fiber given that it is stationary with respect to the family of Killing observers and its refractive index is constant.

Note that it is also possible to introduce an \textit{acceleration potential} $\chi$, given that we assume, as in the Newtonian case, that the Earth is in rigid rotation with constant angular velocity and that no external forces act. \cite{Ehlers1961}
In this case, the acceleration of the congruence of worldlines is $a_\mu = c^2 \partial_\mu \chi$. 
It can be proven that $\chi = \phi$ holds, so acceleration and redshift define the same equipotential surfaces\cite{Salzman1954,philipp2017definition}. 
This is of high practical importance for geodesy, as it means that both acceleration measurements and time/frequency measurements can be used to determine the same geoid in GR. 
The equivalence of the a-geoid and u-geoid has been established before at the 1st-order post-Newtonian level, but now has been shown to hold also in all generality, for example also on the surface of a neutron star given that the assumptions above are met. 

In adapted coordinates $(t,x^1,x^2,x^3)$ where $\exp(\phi)u = \xi = \partial_t$ the general metric reads
\begin{align}
	\label{Eq_StationaryMetric}
	g = e^{2 \phi} \left( -(c dt + \alpha_a dx^a)^2 + \alpha_{ab} dx^a dx^b \right) \, ,
\end{align}
where the latin indices $a, b, \ldots$ run from 1 to 3. The metrical coefficients $\phi$, $\alpha_a$ and $\alpha_{ab}$ are time independent. In these adapted coordinates, ${\exp(2\phi) = -g_{00}}$. 

The redshift and acceleration potentials are defined via specific measurable effects, but as they are equivalent, they are both related to the same \textit{relativistic gravity potential} $U^*$, see Ref.\ \citenum{Philipp2020},
\begin{align}
U^* := c^2 \left(1 - e^{\phi}\right) = c^2 \left(1 - \sqrt{-g_{00}} \right)\, .
\end{align}
Upon expanding $g_{00}$ in a $1/c$ series, in the Newtonian limit we recover the usual gravity potential $W$.
Centrifugal effects are included if the coordinates are adapted to observers that are rigidly corotating with the Earth. 
The relativistic geoid can then be defined as an equipotential surface of the relativistic gravity potential $U^*$. 

We can now reformulate the redshift equation \eqref{Eq_RedshiftPotentialDefinition} in terms of the relativistic gravity potential,
\begin{align}
1 + z & = \frac{\nu_1}{\nu_2} = \frac{1-U^*_2/c^2}{1-U^*_1/c^2}\,,
\end{align}
where $U^*_{1,2}$ refer to the value of the relativistic gravity potential at the events $p$ and $\tilde{p}$ from figure \ref{fig:redshift}, respectively. Note that this result only holds for two clocks of which the worldlines are part of the isometric (Killing) congruence, i.e., clocks that co-rotate rigidly with the Earth. To find an expression at the first post-Newtonian order, we can expand $U^*$ as $U^* = W + \mathcal{O}(c^{-2})$ and finally
\begin{align}
z = \frac{\Delta W}{c^2} + \mathcal{O}(c^{-3})\,.
\end{align}
As the acceleration potential is equivalent to the redshift potential, we can also find an expression for the acceleration $a = -c^2 \mathrm{d}\phi$, i.e., $a_{\mu} = - c^2 \partial_\mu \phi$ in terms of $U^*$,
\begin{align}
a_0 = 0\,, \quad a_i = \frac{-\partial_i U^*}{1-U^*/c^2}\,.
\end{align}
Thus, the acceleration $a_\mu$ is purely spatial. 
The vector field components are obtained by raising the index with the help of the metric.
In the post-Newtonian limit, we then find\cite{Philipp2020}
\begin{subequations}
\begin{align}
    \vec{a}_{\rm pN} & = \nabla U^*_{\rm pN} = \nabla \left( W - \frac{U^2}{2c^2} \right)\,,\\
    a_{\rm pN} &= |\vec{a}_{\rm pN}| \,,
\end{align}
\end{subequations}
recovering the results from Ref.~\citenum{soffel1988relativistic}.

If the two clocks involved in the redshift measurement move arbitrarily, we need to start with Eq.\ \eqref{Eq_RedshiftDefinition2} again and include the respective motion.
We use a metric of the form \eqref{Eq_StationaryMetric}, i.e.,
\begin{align}
    g_{tt} = -c^2 e^{2\phi} \, ,
\end{align}
but assume that the spacetime is static and decompose the redshift into temporal and spatial parts,
\begin{align}
    z = \dfrac{\nu_1 - \nu_2}{\nu_2} = \dfrac{u_1^t(t_1)}{u_2^t(t_2)} \left( \dfrac{1+k_i v_1^i(t_1)}{1+k_i v_2^i(t_2)} \right) -1 =: \dfrac{u_1^t(t_1)}{u_2^t(t_2)} \mathcal{D} -1 \, ,
\end{align}
where the last term is the linear Doppler contribution.
Note that $k_t$ is a constant of motion in any stationary spacetime for a geodesic light signal.
The emission and reception of the light ray happens at times $t_1$ and $t_2$, respectively.
The velocities $v^i$ are measured in the GCRS w.r.t.\ coordinate time TCG, i.e., $u^t v^i = u^i$.
Using the normalization of four-velocities, 
\begin{align}
    g_{\mu\nu} u^\mu u^\nu = (u^t)^2 \left( g_{tt} + g_{ij} v^i v^j \right) = -c^2 \, ,
\end{align}
the redshift is rewritten as
\begin{align}\
    \label{Eq_redshiftGCRS}
    z = \dfrac{\nu_1 - \nu_2}{\nu_2} =: e^{\Delta \phi} \dfrac{\sqrt{1-e^{-2\phi_2}v_2(t_2)^2/c^2}}{\sqrt{1-e^{-2\phi_1}v_1(t_1)^2/c^2}} \mathcal{D} -1 \, .
\end{align}
The clocks' worldlines are parametrized by coordinate time TCG and $\Delta \phi = \phi_2 - \phi_1$, which is the redshift potential difference between the event of emission and the event of reception.
Within the post-Newtonian approximation of GR, in the GCRS we have\cite{Philipp2020} 
\begin{align}
    e^{\phi} = \sqrt{1-\dfrac{2U}{c^2}+\dfrac{2U^2}{c^4}} \, ,
\end{align}
such that the expansion of Eq.\ \eqref{Eq_redshiftGCRS} in powers of $1/c$ yields
\begin{align}
    \label{Eq_redshiftPN}
    z = \dfrac{U_1-U_2}{c^2} + \dfrac{v_1^2}{2c^2} - \dfrac{v_2^2}{2c^2} \, ,
\end{align}
if linear Doppler contributions are neglected. 
Those are
\begin{align}
    \mathcal{D} = \dfrac{1 + \vec{n}\cdot \vec{v}_1/c}{1 + \vec{n}\cdot \vec{v}_2/c} \, ,
\end{align}
but can usually eliminated using two-way ranging and other techniques. 
The vector $\vec{n}$ is the unit vector in the direction of the emitted light ray (line of sight direction).
Note that the emitter-observer problem is implicitly hidden in the redshift equation. 
It consists of finding the null geodesic that connects a chosen emission event on the first worldline with the correct reception event on the second worldline.

\subsubsection{Comparison of relativistic and Newtonian geoids}

The relativistic geoid should now be compared to the conventional Newtonian notion to get an impression of the magnitude of relativistic effects at leading order. 
In Ref.\ \citenum{Philipp2020} it is discussed how to compare the geoid in a first-order post-Newtonian spacetime to the usual Newtonian one. 
In doing so, one has to cope with two subtleties: firstly, one should always be careful to compare results derived in different spacetimes, as coordinates do not have any intrinsic physical meaning in General Relativity and, formally, these concepts are defined on different mathematical manifolds. 
Therefore, it does not make any sense to just directly subtract coordinate values of the geoid surface in a post-Newtonian and a Newtonian spacetime. 
Secondly, there is some ambiguity in the choice of the equipotential surface that we want to call the geoid. 
After all, the value of $W_0$ is just some convention, and so any choice of $U^*_0$ would be possible in principle. 
Thus, this double gauge freedom has to be fixed for a suitable comparison but may very well change the numerical result thereof.

To address the first point, the post-Newtonian geoid, as a two-dimensional surface, can be isometrically embedded into Euclidean space.
Then it is compared with the Newtonian geoid. 
As it turns out, if one directly compares coordinates without using an embedding, the induced error is of the order of the effect!

For the second point, there are intuitive choices for $U^*_0$ that can be considered: i) $U^*_0 = W_0$, (ii) $U^*_0 = W_0 - \frac{U^*_0}{2c^2}$, and (iii) $U^*_0$ such that the resulting relativistic geoid coincides with the Newtonian geoid at a certain reference point. 
Concerning (i), it can be shown that the relativistic geoid is about 2 mm larger than the Newtonian one; for (ii) it is about 4 mm larger; and for (iii) the average difference can be decreased to the $\mu \rm m$ level.

\subsection{Chronometric height}
In geodesy, an important notion of height is the so-called orthometric height.
In fact, it is the formal definition of the common understanding of height above sea level.
However, the information that a chosen location on the Earth's surface is 100$\,$m, say, above sea level does of course not imply that digging into the ground, until this distance is reached, will lead to finding any underground sea surface.
Thus, a chosen mathematical reference, a datum surface, must be constructed. 
In the case of orthometric height this reference height surface is the Earth's conventional geoid, and a global geoid model is necessary to determine the height of any surface location.

For a point $P$ on or above the Earth's surface, the geopotential number 
\begin{align}
    C_P := W_0 - |W_P| 
\end{align}
is defined, where $W_0$ again refers to the value of the gravity potential $W$ on the Newtonian geoid and $W_P$ to its value at the point $P$. The absolute value eliminates conventional, sign-related ambiguities as $W_0$ is always assumed to be positive. If we introduce the average acceleration $\bar{g}$ as
\begin{align}
    \bar{g} & = \frac{1}{H_P} \int_0^{H_P} g \, \mathrm{d}H
\end{align}
where the integration is along the plumbline from $P$ to the geoid, the above equation can be rewritten as
\begin{align}
    C_P = W_0 - |W_P| = H_P \, \dfrac{1}{H_P}\int g \, \mathrm{d}H = H_P \, \bar{g} \quad  \Rightarrow H_P  = \dfrac{C_P}{\bar{g}} \, ,
\end{align}
defining the orthometric height $H_P$.
In the context of chronometry, the difference ${W_0 - |W_P|}$ can be measured with clocks in terms of the gravitational redshift, 
\begin{align}
W_0 - |W_P| = c^2 \, z_P + \mathcal{O}\big( c^3 \big) \, , 
\end{align}
where 
$z_P = e^{(\phi_P-\phi_0)}$ is the redshift between two stationary clocks, rigidly co-rotating with the Earth, positioned on the geoid and at $P$, respectively.
Thus,
\begin{align}
    H_P = \dfrac{c^2 z_P}{\bar{g}} \, .
\end{align}
However, this approach mixes the Newtonian notion of height with first-order relativistic contributions to the gravitational redshift.
To overcome this mixture of concepts, a genuine relativistic height measure can be define via relativistic potential numbers,
\begin{align}\label{eq:relCP}
  C_{P}^{*} := c^2 \left( e^{\phi_P} - e^{\phi_0} \right) = U^*_0 - U^*_P =  c^2 e^{\phi_0} z_P \, ,
\end{align}
which holds at any order in General Relativity.
Here $e^{\phi_0} = 1 - U^*_0/c^2$ is the value of the redshift potential on the relativistic geoid.
To post-Newtonian order, this is
\begin{align}
    C_{P}^{*} = C_P + \dfrac{1}{2c^2} \left( W_0^2 - W_P^2 \right) + \mathcal{O}\big( c^{-3} \big) \, .
\end{align}
We may now proceed to define the genuine relativistic concept of chronometric height.
The chronometric height of a point $P$ is
\begin{align}\label{eq:defHP}
    H_P^* := \dfrac{C_P^*}{\bar{a}} 
    = e^{\phi_0} \dfrac{c^2 z_P}{\bar{a}} \, ,
\end{align}
where $\bar{a}$ is the averaged acceleration $a = c^2 \mathrm{d}\phi$ along the normal w.r.t.\ isochronometric surfaces between $P$ and the geoid

Other height notions such as ellipsoidal heights or normal heights can also be generalized within GR.
The same is true for other reference surfaces and concepts such as the reference ellipsoid and the normal gravity field.

Present efforts aim to realize a global, uniform, and consistent International Height Reference System (IHRS). It is supposed to be a global standard for accurate physical (height) coordinates worldwide. Near the Earth's surface, two stationary standard clocks that are separated only by about 1$\,$cm in height yield a gravitational redshift of about $10^{-18}$, see above. Thus, comparing clocks can in principle lead to accurate height measurements, given high-accuracy clock networks that allow for the observation of pairwise redshifts to obtain the corresponding gravity potential differences. Gravity potential differences between clock sites in a chosen clock network can then lead to an estimate for the transformation parameters between regional or national height reference frames such that distortions are resolved in individual height systems. In practice, the height determination in these networks are to be related to the geopotential numbers. As outlined above, a consistent application of clocks to determine geopotential numbers should make use of the relativistic generalisation in equations \eqref{eq:relCP} and \eqref{eq:defHP}. 

\begin{figure}
    \centering
    \includegraphics[width=0.95\linewidth]{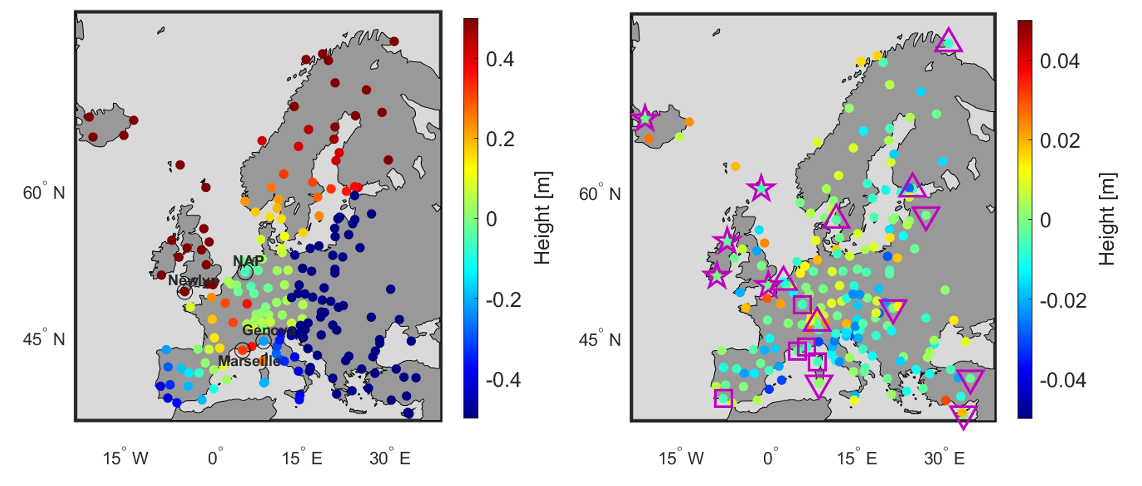}
    \caption{True error before the unification (left) and the residual errors afterwards (right). Different symbols represent the clock sites in each local height system. The error reduced to about $\pm 1\,$cm after the clock-based unification. From Ref.~\citenum{Vincent2024}.}
    \label{fig:IHRS_unification}
\end{figure}

The study in Ref.\ \citenum{Vincent2024}, see also figure \ref{fig:IHRS_unification}, employs chronometric leveling in closed-loop simulations, and the additional value of chronometric measurements is demonstrated for the realization of an IHRS leading to one order of magnitude unification improvement.

\subsection{Full general relativistic degrees of freedom}

In GR, the gravitational field has 10 degrees of freedom instead of only one in Newtonian theory. 
The question is how these degrees of freedom are defined in terms of measurable quantities. Arguably, the most important (in the sense of accessibility by measurements) is the direct generalization of the Newtonian gravity potential above. 

In order to explore the additional degrees of freedom, we first set up a model of the Earth, see, e.g., Refs.~ \citenum{philipp2017definition,Laemmerzahl:2023ejx}. The Earth is given by its constituents, each following a worldline, and these worldlines do not intersect. Although we are interested in mass transport and, thus, temporal changes in the gravitational field, we first start with the stationary situation. This is characterized by the following assumptions: (i) The distances between neighboring constituents are constant. This means that the congruence shows neither shear nor expansion, that is, it is rigid. The Earth can only rotate and accelerate. (ii) The rotation of the Earth is constant. (ii) The acceleration co-rotates with the rotating Earth. This means that the rotation always points towards the same neighboring constituent. From that one concludes that the constituents of the Earth form a Killing congruence. 
That means, the worldline of each constituent is proportional to a Killing vector $\xi$. 

From the mathematical point of view a Killing vector field possesses a norm and a curl
\begin{equation}
e^{2\phi} = g_{\mu\nu} \xi^\mu \xi^\nu \qquad \text{and} \qquad \varpi_\mu =  \epsilon_{\mu\nu}{}^{\rho\sigma} \xi^\rho \partial_\nu \xi^\sigma \, .   
\end{equation}
The gradient $\partial_\mu \phi$  has the interpretation of an acceleration. Thus, as explained in some detail in the foregoing section, $\phi$ is -- up to addition and multiplications with constants -- the general relativistic generalization of the Newtonian potential. The so-called twist $\varpi_\mu$ possesses a potential $\varpi$, $\varpi_\mu = \partial_\mu \varpi$, assuming that Einstein's equations in vacuum hold, that is, outside the Earth. This is in complete analogy to the case where for stationary constellations in electrodynamics there is a scalar potential for the magnetic field. One of the physical effects that can be derived from the twist vector is the Sagnac effect, given by an integration of this vector over a space-like area \cite{Laemmerzahl:2023ejx}. With an extra rotation of the Earth, one has to replace the Killing vector $\xi^\mu$ by $\xi^\mu + \Omega e_\varphi$, where $e_\varphi$ is the unit vector in azimuthal direction. For analytically given space-times, the two potentials can be calculated in an analytic way, see Fig.~\ref{potentials}.

\def\figsubcap#1{\par\noindent\centering\footnotesize(#1)}
\begin{figure}
\begin{center}
\parbox{2.1in}{\includegraphics[width=2in]{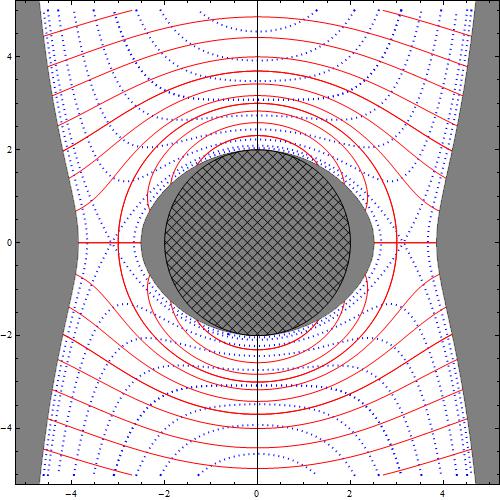}\figsubcap{a}}
  \hspace*{4pt}
  \parbox{2.1in}{\includegraphics[width=2in]{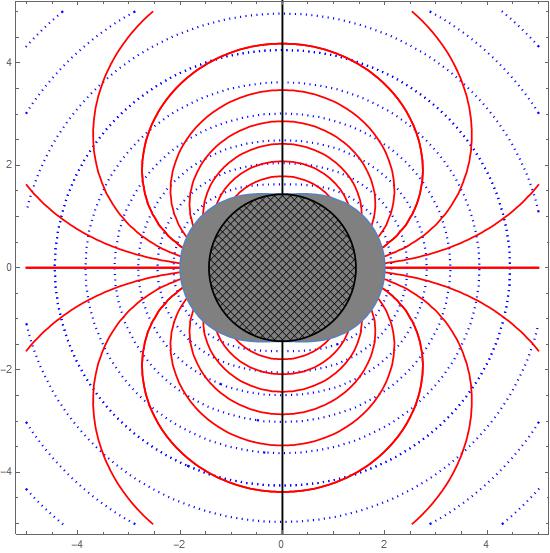}\figsubcap{b}}
\caption{The gravity (blue dotted lines) and twist (red lines) potentials. The crosshatched area is the region inside the black-hole horizon, and the gray shaded area is the region where the potentials are not defined because the rotational velocity becomes larger than the velocity of light. (a) Schwarzschild space-time with additional rotation. (b) Kerr space-time. \label{potentials}}
\end{center}
\end{figure}

In summary, we now introduced two general relativistic potentials: the relativistic gravity potential (sometimes also called gravitoelectric potential) discussed in the foregoing section and the \textit{twist potential} $\varpi$ introduced above (and sometimes called the gravitomagnetic potential). In physical measurements, the acceleration enters with, e.g., falling corner cubes employing classical test bodies while finite potential differences $\Delta\phi$ enter the gravitational redshift of clocks and the phase shift in atom interferometry. The twist vector gives Sagnac-like effects, and is also related to the gravitational Faraday effect and the Schiff effect. 
No effect is known by which finite differences of the gravitomagnetic potential can be measured. 

The line element of a stationary spacetime can be $3 + 1$ decomposed according to \eqref{Eq_StationaryMetric} where we now can add the relation
\begin{equation}
\varpi^a = e^{4 \phi} \epsilon^{abc} \partial_b \alpha_c 
\end{equation}
where $\epsilon^{abc}$ is the volume form associated with the spatial metric $\alpha_{ab}$. Only the components $g_{tt}$ and $g_{ti}$ can be related to potentials. Since we still have the freedom to do a spatial coordinate transformation, we may eliminate three degrees of freedom by requiring, e.g., the spatial metric to be diagonal. As a consequence, we now have five degrees of freedom for the relativistic gravitational field: the two potentials and three components of the spatial metric. 

In a weak field approximation these gravitoelectric and -magnetic terms are the first-order terms. That means that the purely spatial metric is of minor relevance in the context of geodesy. While the gravitoelectric field is related to the energy density of the matter distribution, the gravitomagnetic field is given by the energy current density. Accordingly, the spatial metric is a consequence of the pressure and stress of the matter system. It is clear that the related energy is small and the corresponding gravitational field for the Earth negligible.  

Assuming flat space-time at spatial infinity, for all components of the relativistic space-time metric there exists a series expansion, a multipole expansion taking all the relativistic degrees of freedom into account \cite{SimonBeig1982,Thorne:1980ru}, similar to the expansion \eqref{Eq_NewtonianPotenialDecomposition}. That means, the  practical analysis of data can be carried through in complete analogy to the non-relativistic case. However, there are more degrees of freedom, and one has to discuss for each component of the metric the number of relevant terms for a consistent analysis. Such a scheme with data from GRACE-like missions is currently under development.

Until now, we have assumed a stationary space-time. For the Earth, any time dependence of the gravitational field resulting, e.g., from mass transport, can be included by assuming that the coefficients in the multipole expansion are  time dependent. This is also a common procedure in geodesy. In general, for time-dependent sources of the gravitational field, this is not sufficient. However, the additional terms are related to gravitational waves created by time-dependent sources. For the masses on Earth and their velocities, these parts can be safely neglected. This is a common approximation scheme called an adiabatic approximation. To create measurable gravitational waves, compact objects like Black Holes or Neutron Stars moving at relativistic speeds are required.

\section{Gravity field recovery from space}

In the context of this review, an obvious question is whether clocks in space can be used to determine the gravity field. 
An additional challenge for clocks on satellites is to separate the much larger special relativistic effects, as satellites move with about $10\, \rm km/s$. Fortunately, the usual first-order (or longitudinal) Doppler effect can be eliminated by using a two-way link \cite{giuliani_determination_2024}. 
However, measurement uncertainties in the second-order (or transversal) Doppler effect map into the gravitational potential determination. 
In Ref. \citenum{muller_using_2020} a rough estimate was given: determining the geoid with a precision of about 1 cm corresponds to measuring the velocity of a satellite in a Low Earth Orbit (LEO) with a precision of about $10\, \rm \mu m/$s, which seems to be extremely challenging.

In our investigation of the capabilities of clock measurements in gravity field recovery (GFR) from space, we utilize a closed-loop approach. 
It consists of a high precision orbit simulation using the eXtended Hybrid simulation Platform for Space systems (XHPS) \cite{woske_grace_2019, huckfeldt_grace_2024}, in which realistic mock data is generated, by adding variable noise to the simulation results. 
The mock data is then used in a gravity field recovery process, applying the state-of-the-art variational-equation approach. The loop is closed by comparison of the estimated gravity field with the model used as input for the orbit simulation. 
With this approach, not only different types of measurements but also environmental effects, orbit designs, satellite constellations, noise models, etc.\ can be investigated. 

The XHPS is based on the Newtonian representation of the gravitational potential, which can be extended by so-called post-Newtoninan Schwarzschild, Lense-Thirring and de Sitter corrections \cite{petit_iers_2010}. 
It's unique capability to model the external non-gravitational accelerations acting on satellites as well as realistic attitude control and other features of a spacecraft system, make the XHPS the best suited tool for the generation of high-precision mock data. 
By adding noise to the "perfect" simulation results, arbitrary measurements can be generated. For example, processing position and velocity data yields GNSS measurements. Furthermore, the inter-satellite ranging data of the GRACE missions can be generated including the characteristics of the sensor \cite{woske_grace_2019}.

For investigating clocks, the capability to simulate the frequency variation and propagation of the proper time of a clock on a satellite has been added to the XHPS. 
This feature has already been used in the exploration of the gravitomagnetic clock effect of clocks on the GALILEO satellites \cite{scheumann_gravitomagnetic_2023}. 
Similarly, the GFR tool has to be extended to handle clock measurements. 
The variational-equation approach of the GFR tool has the advantage that not only the multipole coefficients of the gravitational field are estimated but also the satellite's initial states, being the position and velocity. 

The first tests are conducted with a simplified clock measurement that includes the following assumptions: The propagation time of the laser is neglected as well as the transversal Doppler effect. 
It is also assumed that the clock comparison can be done, without any conditions for ground station contact times. 
Therefore, the observation equation describing the fractional frequency shift is Eq.\ \eqref{Eq_redshiftPN},
\begin{align}
    z = \frac{\Delta \nu}{\nu_2} = \frac{U_1 - U_2}{c^2} + \frac{v_1^2 - v_2^2}{2c^2},
\end{align}
where the subscript "2" denotes properties of the receiver and "1" properties of the transmitter. Furthermore, $U$ is the gravitational potential at receiver/transmitter positions, $v$ the speed of the receiver/transmitter in the GCRS and $c$ the speed of light.

\begin{figure}
\centering
\parbox{2.1in}{\includegraphics[width=2in]{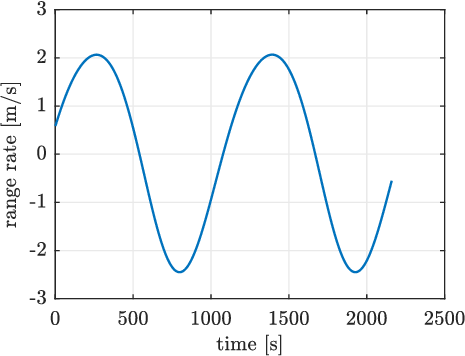}\figsubcap{a}}
  \hspace*{4pt}
  \parbox{2.1in}{\includegraphics[width=2in]{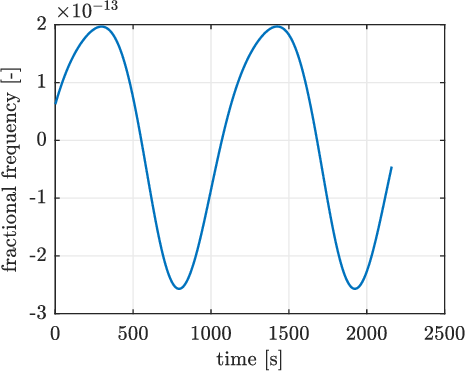}\figsubcap{b}}
\caption{Simulated microwave ranging and clock measurements for a GRACE-like scenario. (a) Range rate. (b) Inter-satellite clock comparison.}
\label{fig:rangerateVSintersatqcl}
\end{figure}

\begin{figure}
\centering
\includegraphics[width=0.5\textwidth]{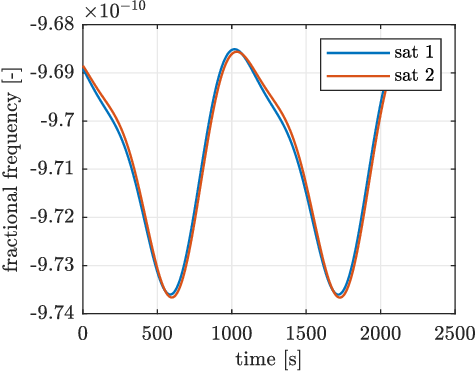}
\caption{Satellite-geoid clock comparison}
\label{fig:satgeoidqcl}
\end{figure}

To evaluate the quality of the clock measurements in the GFR processing, comparisons of results for a GRACE-like scenario have been conducted. In this constellation the second satellite is assumed to follow the leading satellite on the same trajectory with a distance of about 200 km. Within such a scenario, the new measurement scheme can be assessed in comparison to the currently flying and commonly used observation type in GFR. The first clock comparison setup is the fractional frequency measurement of clocks on two satellites and has very similar characteristics to the range-rate measurement of the GRACE microwave ranging sensor. Figure\,\ref{fig:rangerateVSintersatqcl} shows the two simulated observations for the same satellite orbits over a three hour period. As the second setup a measurement between clocks on the satellites and a clock on the geoid was implemented. The clock on the geoid is assumed to be at rest. In combination with the above mentioned assumptions, the observations are continuous and the signal can be increased by three orders of magnitude due to the increased height and velocity differences, compare Figures\,\ref{fig:rangerateVSintersatqcl}(b) and \ref{fig:satgeoidqcl}.

\begin{figure}
\centering
\includegraphics[width=1\textwidth]{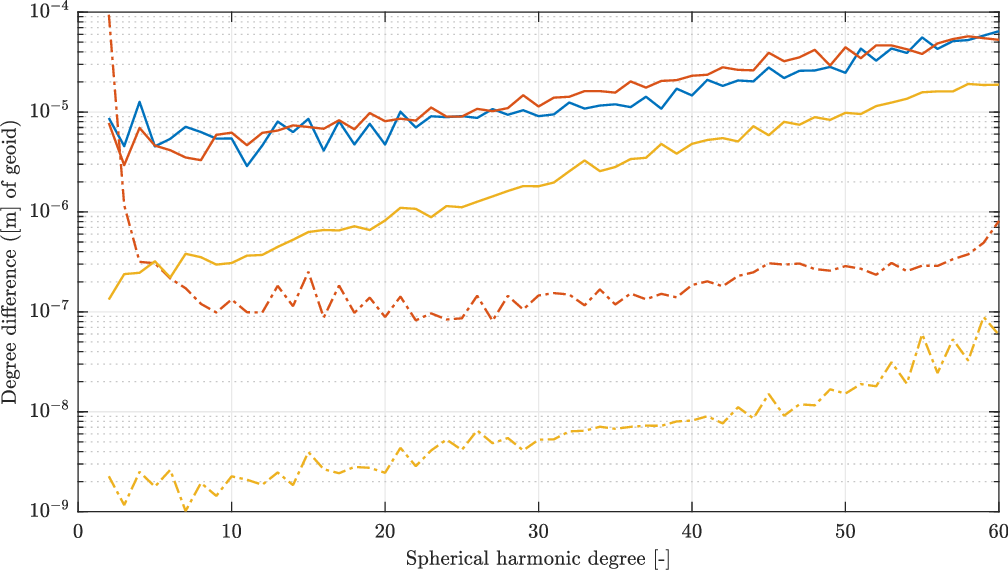} \\
\includegraphics[scale=0.9]{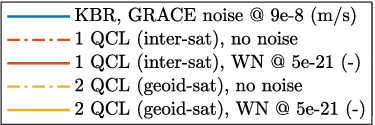}
\caption{Degree differences of multiple GFR results with various noise levels. Here QCL refers to Quantum CLock and the digit indicates the number of pairwise clock comparisons.}
\label{fig:gfrresults}
\end{figure}

The results of the GFR process are shown in Figure\,\ref{fig:gfrresults} as degree differences as described in Ref. \citenum{woske_gravity_2021},
\begin{align}\label{eq:defdn}
    \Delta d_n = R\sqrt{\sum_{m=0}^{n}\left( \Delta C_{nm}^2 + \Delta S_{nm}^2 \right)}\,,
\end{align}
where $R$ denotes the reference radius of Earth and $C_{nm}, S_{nm}$ are the spherical harmonic coefficients of order $n$ and degree $m$ of the gravitational field. The delta operator indicates that differences of two gravitational fields are evaluated. With this representation we can assess how well the estimated gravitational field matches the reference field, which was used for the data simulation. By including the reference radius $R$ in Eq. \eqref{eq:defdn}, the degree differences can be understood as errors of the geoid height in meters.

In blue, the results with the GRACE microwave (K-Band) Ranging sensor (KBR) are shown, which include a noise model that has been derived from the actual sensor and has a standard deviation $\sigma = 9\cdot10^{-8} \rm m/s$. In yellow and red, results with the two different clock setups are pictured. Dashed lines indicate results where no noise was added to the observations and solid lines results with observations including a white noise. In the inter-satellite measurement setup one clock observation is used, whereas in the geoid-satellite setup this number is increased to two due to a clock comparison for each satellite with a clock on the geoid.

With perfect clock observations both measurement setups unsurprisingly perform better than the reference KBR measurement. However, adding noise with a standard deviation as small as $5\cdot10^{-21}$ to the clock observations lead to degree differences closely matching the reference case. This is arguably due to the influence of the special relativistic time dilation term and the satellite constellation. Because both satellites are in leader-follower constellation the potential difference $\Delta U$, which contains the desired multipole coefficients, is very small and the estimation process is impeded. By increasing the potential difference through clock comparison between clocks on the satellites and the geoid, the results can be improved.

An uncertainty of a stationary clock in the range of $10^{-21}$ is currently impossible, much less so for a moving clock on a satellite. Additionally, the noise model does not account for the time variability of the clock uncertainty. The conducted investigation therefore only shows the technical feasibility of clock measurements for the estimation of Earth's gravity field in general. 

The variational equation approach uses only satellite positions as observations and as stated before co-estimates the satellites initial position and velocity during the estimation process. Therefore, the velocity precision requirement of $10\, \rm \mu m/s$ as estimated in  Ref.~\citenum{muller_using_2020} does not apply here, and below $\rm cm$-level geoid precision is mostly limited by clock precision. 

Currently, development of a more realistic clock measurement including the transversal Doppler effect, realistic noise models, propagation time of the signal as well as Doppler-canceling techniques is in progress, to allow more realistic closed-loop simulations of clock measurements in GFR. Future investigations will also include the design of best suited satellite constellations, including high-low measurement options and multi-satellite constellations.

\section{Discussion and outlook}
In this paper, we have explored the influence of General Relativity (GR) on modern geodesy, illustrating its role in both refining classical measurement techniques and pioneering new methodologies through the gravitational redshift. We reviewed the construction of the relativistic gravity potential and the corresponding geoid definition as an equipotential surface developed in Refs \refcite{philipp2017definition,Philipp2020}, and demonstrated the conceptual and practical differences from the Newtonian geoid. Although the difference is currently below the measurement accuracies, clocks on the $10^{-19}$ level will become available in the near future. Therefore, further discussions are warranted on the question how to select the "relativistic mean sea level" $U^*_0$, which might have significant impact, e.g. on the definition of the international atomic timescale TAI or the definition of an International Height Reference System in the future. The IHRS will be based on geopotential numbers, and the benefit of using clocks for unifying height systems has already been demonstrated in simulations. Therefore, a consistent scheme including a relativistic definition of geopotential numbers like the one introduced in this paper should be employed.

Additionally, we discussed the construction of potentials related to the non-Newtonian degrees of freedom of the relativistic gravitational field. We discussed the twist potential in some detail, which is related to rotational effects like the Sagnac effect. An intriguing application might be the construction of a reference frame based on the relativistic gravity potential and the twist potential. The practical use of such a reference system in geodesy, and of the twist potential in general,  is unclear as of right now, and warrants further investigations. The role of the additional degrees of freedom of the spatial parts of the metric in geodesy is even more elusive, and currently and in the near future probably of negligible importance. 

Finally, we evaluated the potential of clock-based observations for Gravity Field Recovery (GFR) from space. These first results show a general benefit of clocks in satellite geodesy to increase the accuracy of gravity fields. A favorable setup for clock measurements includes different orbit heights of the clocks to increase the gravitational redshift part in the measurement. Especially the lower coefficients can potentially be estimated more precisely than with KBR. Comparison with KBR observations also showed that the clock noise level is a major bottleneck in the estimation procedure, and currently unobtainable levels are necessary in the leader-follower constellation to see any benefit of clocks. However, with the right estimation procedure, previously stated requirements of the velocity data accuracy do not apply.

\section*{Acknowledgments}

We acknowledge the support by the Deutsche Forschungsgemeinschaft (DFG, Germany Research Foundation) – Project-ID 434617780 – SFB 1464 and under
Germany’s Excellence Strategy – EXC-2123 QuantumFrontiers – 390837967.

\bibliographystyle{ws-procs961x669}
\bibliography{sample}

\begin{thebibliography}{10}

\bibitem{vermeer_chronometric_1983}
M.~Vermeer, {\em Chronometric Levelling}Reports of the Finnish Geodetic
  Institute, Reports of the Finnish Geodetic Institute (Geodeettinen Laitos,
  Geodetiska Institutet.

\bibitem{Bjerhammar1985}
A.~Bjerhammar, On a relativistic geodesy, {\em Bulletin g{\'e}od{\'e}sique}
  {\bf 59}, 207  (1985).

\bibitem{Herbers:22}
S.~Herbers, S.~H{\"a}fner, S.~D{\"o}rscher, T.~L{\"u}cke, U.~Sterr and
  C.~Lisdat, Transportable clock laser system with an instability of 1.6$\times
  10^{-16}$, {\em Optics Letters} {\bf 47}, 5441  (2022).

\bibitem{Lisdat2021}
C.~Lisdat, S.~D\"orscher, I.~Nosske and U.~Sterr, Blackbody radiation shift in
  strontium lattice clocks revisited, {\em Phys. Rev. Res.} {\bf 3}, p. L042036
  (Dec 2021).

\bibitem{Doerscher2023}
S.~D\"orscher, J.~Klose, S.~Maratha~Palli and C.~Lisdat, Experimental
  determination of the $e2\text{\ensuremath{-}}m1$ polarizability of the
  strontium clock transition, {\em Phys. Rev. Res.} {\bf 5}, p. L012013 (Feb
  2023).

\bibitem{Schioppo2022}
M.~Schioppo, J.~Kronjaeger, A.~Silva, R.~Ilieva, J.~Paterson, C.~Baynham,
  W.~Bowden, I.~Hill, R.~Hobson, A.~Vianello, M.~Dovale~Álvarez, R.~Williams,
  G.~Marra, H.~Margolis, A.~Amy-Klein, O.~Lopez, E.~Cantin, H.~Álvarez
  Martínez, R.~Targat and G.~Grosche, Comparing ultrastable lasers at 7
  $\times 10^{-17}$ fractional frequency instability through a 2220 km optical
  fibre network, {\em Nature Communications} {\bf 13} (01 2022).

\bibitem{Koke:2019}
S.~Koke, A.~Kuhl, T.~Waterholter, S.~M.~F. Raupach, O.~Lopez, E.~Cantin,
  N.~Quintin, A.~Amy-Klein, P.-E. Pottie and G.~Grosche, {Combining fiber
  Brillouin amplification with a repeater laser station for fiber-based optical
  frequency dissemination over 1400 km}, {\em New J. Phys.} {\bf 21}, p. 123017
   (2019).

\bibitem{Elmaghraby2023}
A.~Elmaghraby, T.~Krawinkel, S.~Schön, D.~Piester and A.~Bauch, On error
  modeling in gnss-based frequency transfer: Effects of temperature variations
  and satellite orbit repeat times02 2023.

\bibitem{Takamoto2020}
M.~{Takamoto}, I.~{Ushijima}, N.~{Ohmae}, T.~{Yahagi}, K.~{Kokado},
  H.~{Shinkai} and H.~{Katori}, {Test of general relativity by a pair of
  transportable optical lattice clocks}, {\em Nature Photonics} {\bf 14}, 411
  (April 2020).

\bibitem{Grotti2024}
J.~Grotti, I.~Nosske, S.~Koller, S.~Herbers, H.~Denker, L.~Timmen,
  G.~Vishnyakova, G.~Grosche, T.~Waterholter, A.~Kuhl, S.~Koke, E.~Benkler,
  M.~Giunta, L.~Maisenbacher, A.~Matveev, S.~D\"orscher, R.~Schwarz,
  A.~Al-Masoudi, T.~H\"ansch, T.~Udem, R.~Holzwarth and C.~Lisdat,
  Long-distance chronometric leveling with a portable optical clock, {\em Phys.
  Rev. Appl.} {\bf 21}, p. L061001 (Jun 2024).

\bibitem{muller2008geodesy}
J.~M{\"u}ller, M.~Soffel and S.~A. Klioner, Geodesy and relativity, {\em
  Journal of Geodesy} {\bf 82}, 133  (2008).

\bibitem{Soffel:2003cr}
M.~Soffel {\em et~al.}, {The IAU 2000 resolutions for astrometry, celestial
  mechanics and metrology in the relativistic framework: Explanatory
  supplement}, {\em Astron. J.} {\bf 126}, 2687  (2003).

\bibitem{TorgeMueller+2012}
W.~Torge and J.~M{\"u}ller, {\em Geodesy} (De Gruyter, Berlin, Boston, 2012).

\bibitem{sanchez2016conventional}
L.~S{\'a}nchez, R.~{\v{C}}underl{\'\i}k, N.~Dayoub, K.~Mikula,
  Z.~Minarechov{\'a}, Z.~{\v{S}}{\'\i}ma, V.~Vatrt and
  M.~Vojt{\'\i}{\v{s}}kov{\'a}, A conventional value for the geoid reference
  potential w\_ 0 w 0, {\em Journal of Geodesy} {\bf 90}, 815  (2016).

\bibitem{soffel1988relativistic}
M.~Soffel, H.~Herold, H.~Ruder and M.~Schneider, Relativistic theory of
  gravimetric measurements and definition of the geoid, {\em manuscripta
  geodaetica} {\bf 13}, 143  (1988).

\bibitem{kopeikin2015towards}
S.~M. Kopeikin, E.~M. Mazurova and A.~P. Karpik, Towards an exact relativistic
  theory of earth's geoid undulation, {\em Physics Letters A} {\bf 379}, 1555
  (2015).

\bibitem{oltean2016geoids}
M.~Oltean, R.~J. Epp, P.~L. McGrath and R.~B. Mann, Geoids in general
  relativity: geoid quasilocal frames, {\em Classical and Quantum Gravity} {\bf
  33}, p. 105001  (2016).

\bibitem{philipp2017definition}
D.~Philipp, V.~Perlick, D.~Puetzfeld, E.~Hackmann and C.~L{\"a}mmerzahl,
  Definition of the relativistic geoid in terms of isochronometric surfaces,
  {\em Physical Review D} {\bf 95}, p. 104037  (2017).

\bibitem{Kermack1934}
W.~O. Kermack, W.~H. McCrea and E.~T. Whittaker, Iv.---on properties of null
  geodesics, and their application to the theory of radiation., {\em
  Proceedings of the Royal Society of Edinburgh} {\bf 53}, 31 (1 1934).

\bibitem{HassePerlick1988}
W.~Hasse and V.~Perlick, Geometrical and kinematical characterization of
  parallax-free world models, {\em Journal of Mathematical Physics} {\bf 29},
  2064  (1988).

\bibitem{Ehlers1961}
J.~{Ehlers}, {Beitr{\"a}ge zur relativistischen Mechanik kontinuierlicher
  Medien}, {\em Mainz Akademie Wissenschaften Mathematisch
  Naturwissenschaftliche Klasse} {\bf 11}  (1961).

\bibitem{Salzman1954}
G.~Salzman and A.~H. Taub, Born-type rigid motion in relativity, {\em Phys.
  Rev.} {\bf 95}, 1659 (Sep 1954).

\bibitem{Philipp2020}
D.~Philipp, E.~Hackmann, C.~L\"ammerzahl and J.~M\"uller, Relativistic geoid:
  Gravity potential and relativistic effects, {\em Phys. Rev. D} {\bf 101}, p.
  064032 (Mar 2020).

\bibitem{Vincent2024}
A.~Vincent, J.~Müller, C.~Lisdat and D.~Philipp, Realization of a clock-based
  global height system: A simulation study for europe and south america
  (2024), arXiv:2411.07888 [gr-qc].

\bibitem{Laemmerzahl:2023ejx}
C.~L{\"a}mmerzahl and V.~Perlick, {Potentials for general-relativistic
  geodesy}, {\em Phys. Rev. D} {\bf 109}, p. 044028  (2024).

\bibitem{SimonBeig1982}
W.~Simon and R.~Beig, {The multipole structure of stationary space-times}, {\em
  J. Math. Phys.} {\bf 24}, p. 1163  (1982).

\bibitem{Thorne:1980ru}
K.~S. Thorne, {Multipole Expansions of Gravitational Radiation}, {\em Rev. Mod.
  Phys.} {\bf 52}, 299  (1980).

\bibitem{giuliani_determination_2024}
S.~Giuliani, B.~D. Tapley and J.~C. Ries, Determination of the time-variable
  geopotential by means of orbiting clocks, {\em Journal of Geodesy} {\bf 98},
  p.~50 (June 2024).

\bibitem{muller_using_2020}
J.~Müller and H.~Wu, Using quantum optical sensors for determining the
  {Earth}’s gravity field from space, {\em Journal of Geodesy} {\bf 94},
  p.~71 (August 2020).

\bibitem{woske_grace_2019}
F.~Wöske, T.~Kato, B.~Rievers and M.~List, {GRACE} accelerometer calibration
  by high precision non-gravitational force modeling, {\em Advances in Space
  Research} {\bf 63}, 1318 (February 2019).

\bibitem{huckfeldt_grace_2024}
M.~Huckfeldt, F.~Wöske, B.~Rievers and M.~List, {GRACE} {Follow}-{On}
  accelerometer data recovery by high-precision environment modelling, {\em
  Advances in Space Research} {\bf 73}, 5783 (June 2024).

\bibitem{petit_iers_2010}
G.~Petit and B.~Luzum, {IERS} {Technical} {Note} {No}. 36  (2010).

\bibitem{scheumann_gravitomagnetic_2023}
J.~Scheumann, D.~Philipp, S.~Herrmann, E.~Hackmann, B.~Rievers,
  J.~Ventura-Traveset, L.~Mendes and C.~Lämmerzahl, Gravitomagnetic {Clock}
  {Effect}: {Using} {GALILEO} to explore {General} {Relativity} (November
  2023), arXiv:2311.12018 [gr-qc].

\bibitem{woske_gravity_2021}
F.~Wöske, Gravity {Field} {Recovery} from {GRACE} {Satellite} {Data} and
  {Investigation} of {Sensor}, {Environment} and {Processing}-{Option}
  {Influences} by {Closed} {Loop} {Mission} {Simulation}, PhD thesis,
  Universität BremenOctober 2021.

\end{thebibliography}

\end{document}